\documentclass[]{aa}  
\usepackage{graphicx}
\usepackage{txfonts}

\usepackage{epsf}
\usepackage{natbib}
\usepackage{astro_bib_macro}
\usepackage{amssymb}
\bibpunct{(}{)}{;}{a}{}{,}

\newcommand{\kms}{km\,s$^{-1}$}
\newcommand{\fuse}{\emph{FUSE}}
\newcommand{\visir}{\emph{VISIR}}
\newcommand{\texes}{\emph{TEXES}}
\newcommand{\uves}{\emph{UVES}}

\newcommand{\vrad}{v$_{\rm rad}$} 
\newcommand{\flux}{ergs\,s$^{-1}$\,cm$^{-2}$}

\newcommand{\owens}{{\sc Owens}}
\newcommand{\Av}{A$_{\rm v}$}

\begin{document}
\title{Where is the warm H$_2$ ? \\
  A search for H$_2$ emission from disks around Herbig Ae/Be stars}

   \author{C. Martin-Za{\"\i}di\inst{1}  
     \and 
     J-.C. Augereau\inst{1}
     \and
     F. M\'enard\inst{1}
     \and 
     J. Olofsson\inst{1} 
     \and
     A. Carmona\inst{2}
     \and
     C. Pinte\inst{3}
     \and
     E. Habart\inst{4}  }

   \offprints{C. Martin-Za{\"\i}di}

   \institute{Universit\'e Joseph Fourier - Grenoble 1 / CNRS,
     Laboratoire d'Astrophysique de Grenoble (LAOG) UMR 5571, BP 53,
     38041 Grenoble Cedex 09, France \\
     \email{claire.martin-zaidi@obs.ujf-grenoble.fr} 
     \and
     ISDC Data Centre for Astrophysics \& Geneva Observatory,
     University of Geneva, chemin d'Ecogia 16, 1290 Versoix,
     Switzerland
     \and 
     School of Physics, University of Exeter, Stocker Road, Exeter EX4
     4QL, United Kingdom
     \and 
     Institut d'Astrophysique Spatiale, Universit\'e Paris-Sud, 91405
     Orsay Cedex, France}

   \date{Received ... / Accepted ...}

  \abstract
  {Mid-infrared (mid-IR) emission lines of molecular hydrogen (H$_2$)
  are useful probes to determine the mass of warm gas present in the
  surface layers of circumstellar disks. In the past years, numerous
  observations of Herbig Ae/Be stars (HAeBes) have been performed, but
  only two detections of H$_2$ mid-IR emission toward HD~97048 and
  AB~Aur have been reported.}
{ We aim at tracing the warm gas in the circumstellar environment of
  five additional HAeBes with gas-rich environments and/or physical
  characteristics close to those of AB~Aur and/or HD~97048, to discuss
  whether the detections toward these two objects are suggestive of
  peculiar conditions for the observed gas.}
{ We search for the H$_2$ S(1) emission line at 17.035 $\mu$m using
  high-resolution mid-IR spectra obtained with VLT/\visir, and
  complemented by CH molecule observations with VLT/\uves. We gather
  the H$_2$ measurements from the literature to put the new results in
  context and search for a correlation with some disk properties.}
{ None of the five \visir\ targets shows evidence for H$_2$ emission
  at 17.035 $\mu$m. From the 3$\sigma$ upper limits on the integrated
  line fluxes we constrain the amount of optically thin warm
  ($>150~K$) gas to be less than $\sim 1.4~ M_{\rm Jup}$ in the disk
  surface layers. There are now 20 HAeBes observed with \visir\ and
  \texes\ instruments to search for warm H$_2$, but only two
  detections (HD~97048 and AB~Aur) were made so far. We find that the
  two stars with detected warm H$_2$ show at the same time high
  30/13\,$\mu$m flux ratios and large PAH line fluxes at 8.6 and
  11.3$\,\mu$m compared to the bulk of observed HAeBes and have
  emission CO lines detected at $4.7\,\mu$m.  We detect the CH
  4300.3\,$\AA$ absorption line toward both HD~97048 and AB~Aur with
  \uves. The CH to H$_2$ abundance ratios that this would imply if it
  were to arise from the same component as well as the radial velocity
  of the CH lines both suggest that CH arises from a surrounding
  envelope, while the detected H$_2$ would reside in the disk.}
{The two detections of the S(1) line in the disks of HD~97048 and
  AB~Aur suggest either peculiar physical conditions or a particular
  stage of evolution. New instruments such as \textit{Herschel / PACS}
  should bring significant new data for the constraints of
  thermodynamics in young disks by observing the gas and the dust
  simultaneously.}


   \titlerunning{Where is the warm H$_2$ ?}
   \authorrunning{C. Martin-Za{\"\i}di et al.}

   \maketitle
%

\section{Introduction}

Planets are supposed to form in circumstellar disks composed of gas
and dust around stars in their pre-main sequence phase. At this
evolutionary stage, the disk mass is essentially dominated by the gas
(99\%), especially molecular hydrogen (H$_2$). Although H$_2$ is the
principal gaseous constituent in disks, it is very challenging to
detect. Molecular hydrogen is an homonuclear molecule which means that
its fundamental transitions are quadrupolar in nature. Hence their
Einstein spontaneous emission coefficients are small and produce only
weak lines. For circumstellar disks another challenge is that the weak
H$_2$ lines must be detected on top of the strong dust continuum
emission. In the mid-infrared, high spectral resolution instruments
like \visir\ at the VLT are required to disentangle these weak lines
from the continuum.

However, one should bear in mind a fundamental issue concerning the
structure of a gas-rich optically thick disk. Molecular lines are
produced in the hot upper surfaces of the disk where the molecular gas
and accompanying dust are optically thin. Therefore, molecular line
emission is not sensitive to and does not probe the mid-plane interior
layers of the disk because it is optically thick. Because the amount
of molecular gas in the optically thin surface layers is small, the
expected H$_2$ line fluxes are very weak. As an example of this,
\cite{Carmona08} calculated the expected H$_2$ line fluxes from
typical Herbig Ae disks assuming a two-layer \cite{Chiang97} disk
model, $T_{\rm gas}=T_{\rm dust}$, a gas-to-dust-ratio of 100, LTE
emission and a distance of 140 pc. They found that the expected line
H$_2$ fluxes are much fainter than the detection limits of current
instrumentation. Indeed, numerous non-detections of H$_2$ mid-IR pure
rotational lines in the circumstellar environment of young stars have
been reported in the past few years \citep{Bitner08, Carmona08,
  klr08b, klr09a}. Nevertheless, lines can reach a detectable level if
the gas-to-dust to ratio is allowed to be higher than 100 and/or the
$T_{\rm gas}>T_{\rm dust}$ in the surface layers of the disk.

%
\begin{table*}
\begin{center}
  \caption{Astrophysical parameters of the sample stars.}
\begin{tabular}{lcccccccccccccccccc}
  \hline
  \hline
  Star        & Sp.    & $T_{eff}$     & \Av         &  \vrad $^{(a)}$  & $d$        & Disk         & Evidence for  \\
              & Type   & (K)          & (mag)        & (\kms)       & (pc)       & resolved     &  gas-rich \\
              &         &              &              &              &            &              &  environment \\
  \hline
  HD~142527   & F6~IIIe & 6300$^{(1)}$  & 1.49$^{(1)}$ & -3.5$^{(2)}$  & 140$^{(3)}$ & yes$^{(4)}$   &   \\
  HD~169142   & A8~Ve   & 8130$^{(5)}$  & 0.37$^{(6)}$ & -3.0$^{(7)}$  & 145$^{(5)}$ & yes$^{(8)}$   & yes$^{(9)}$   \\
  HD~150193A  & A1~Ve   & 9300$^{(1)}$  & 1.61$^{(1)}$ & -6.0$^{(10)}$  & 150$^{(1)}$ & yes$^{(11)}$  &    \\
  HD~163296   & A1~Ve   & 9300$^{(1)}$  & 0.25$^{(1)}$ & -4.0$^{(2)}$  & 122$^{(1)}$ & yes$^{(12)}$  & yes$^{(13)}$    \\
  HD~100546   & B9~Vne  & 10470$^{(1)}$ & 0.25$^{(1)}$ & +17$^{(13)}$   & 103$^{(1)}$ & yes$^{(14)}$ & yes$^{(13)}$    \\
  \hline
\end{tabular}
\begin{list}{}{} 
\item $^{(a)}$ radial velocity of the star in the heliocentric rest
  frame.

\item References: $^{(1)}$ \cite{VdA98b}; $^{(2)}$ SIMBAD database;
  $^{(3)}$ \cite{DeZeeuw_99}; $^{(4)}$ \cite{Fukagawa06}; $^{(5)}$
  \cite{Acke05}; $^{(6)}$ \cite{Malfait98}; $^{(7)}$ \cite{Dunkin97};
  $^{(8)}$\cite{Kuhn01}; $^{(9)}$ \cite{Panic08}; $^{(10)}$
  \cite{Reipurth96}; $^{(11)}$ \cite{Fukagawa03}; $^{(12)}$
  \cite{GRADY00}; $^{(13)}$ \cite{LECAV03}; $^{(14)}$ \cite{Augereau01}. \\
\end{list}
\label{param}
\end{center}
\end{table*}

\begin{table*}
\begin{center}
  \caption{Summary of the observations.}
\begin{tabular}{lcccccccccccccccccc}
  \hline
  \hline
  Star      &  $t_{exp}$ & Airmass   & Optical   & Standard  & Airmass   & Optical   & Asteroid  & Airmass   & Optical  \\
            &  (s)      &           & Seeing    &  Star     &           & Seeing    &           &           & Seeing     \\
            &           &           & ('')      &           &           &  ('')     &           &           &  ('')    \\
  \hline
HD~142527   & 3600      & 1.10-1.28 & 0.71-1.08 & HD~211416 & 1.23-1.25 & 0.79-1.02 & HEBE      & 1.19-1.20 & 0.97-1.06 \\
HD~169142   & 3600      & 1.17-1.50 & 0.71-1.01 & HD~211416 & 1.23-1.25 & 0.79-1.02 & HEBE      & 1.19-1.20 & 0.97-1.06 \\
HD~150193A  & 3600      & 1.31-1.88 & 0.85-1.40 & HD~211416 & 1.24-1.25 & 0.81-0.99 & PSYCHE    & 1.13-1.16 & 0.75-0.83 \\
HD~163296   & 3600      & 1.60-2.70 & 0.79-1.19 & HD~211416 & 1.24-1.25 & 0.81-0.99 & IRIS      & 1.00-1.01 & 0.99-1.28 \\
HD~100546   & 3600      & 1.44-1.52 & 0.73-1.09 & HD~89388  & 1.31-1.32 & 0.61-0.73 & HEBE      & 1.19-1.20 & 0.97-1.06 \\
  \hline
\end{tabular}
\begin{list}{}{} 
\item  The airmass and seeing intervals are given from the beginning to the
  end of the observations. \\
\end{list}

\label{log_obs}
\end{center}
\end{table*}

Molecular hydrogen mid-IR lines have been detected in two Herbig Ae/Be
stars, namely HD~97048 and AB Aur, respectively observed with
VLT/\visir\ \citep{klr07} and \texes\ \citep{Bitner07}. These
detections imply particular physical conditions for the gas and dust,
such as $T_{\rm gas}>T_{\rm dust}$, as mentioned above. These
conditions may be created by gas heated by X-rays or UV photons in the
surface layers of the disks. We note that these two detections
indicate that the gas has not completely dissipated in the inner part
of these disks in a lifetime of about 3 Myrs (ages of the stars),
while photoevaporation of the gas is expected to clear up this inner
region within roughly the same time \citep[$<$3 Myrs;
e.g.][]{Takeuchi05, Alexander08}. Indeed, \cite{klr09a} have shown
that the emitting H$_2$ around HD~97048 was more likely distributed in
an extended region within the inner disk, between 5~AU and 35~AU of
the disk, and \cite{Bitner07} concluded from their observation of the
disk of AB~Aur that the H$_2$ emission arised around 18~AU from the
central star.

We present new observations with the \visir\ high-resolution
spectroscopic mode, to search for the mid-IR H$_2$ emission line at
17.035 $\mu$m, the most intense pure rotational line observable from
the ground, in a sample of well studied Herbig Ae/Be stars known to
harbor extended gas-rich circumstellar disks. The main goal of these
observations is to enlarge the global sample of HAeBes observed at
17.035 $\mu$m, and to better constrain the particular physical
conditions observed in the disks of HD~97048 and AB~Aur. In order to
better understand the detections of H$_2$ in the disks of HD~97048 and
AB~Aur, we also analyze VLT/\uves\ spectra of these two stars to
observe spectral lines of the CH and CH$^+$ molecules which are linked
to the formation and excitation of H$_2$. In Sects.~\ref{results} and
\ref{discuss} we present the analysis of the \visir\ spectra of the
five HAeBes of our sample and relate it to the context of the global
search for molecular hydrogen at mid-IR wavelengths. We perfom a
statistical analysis of the whole sample of HAeBes where H$_2$
emission has been searched for in the mid-IR, and explore the possible
link between the H$_2$ and the dusty disk properties. Finally, we
discuss the possible origin of the mid-IR H$_2$ emission in the disks
of HD~97048 and AB~Aur.

\section{Observation and data reduction}
\label{analysis}

\begin{figure*}[!htbp]
\begin{center}
\includegraphics[width=17cm]{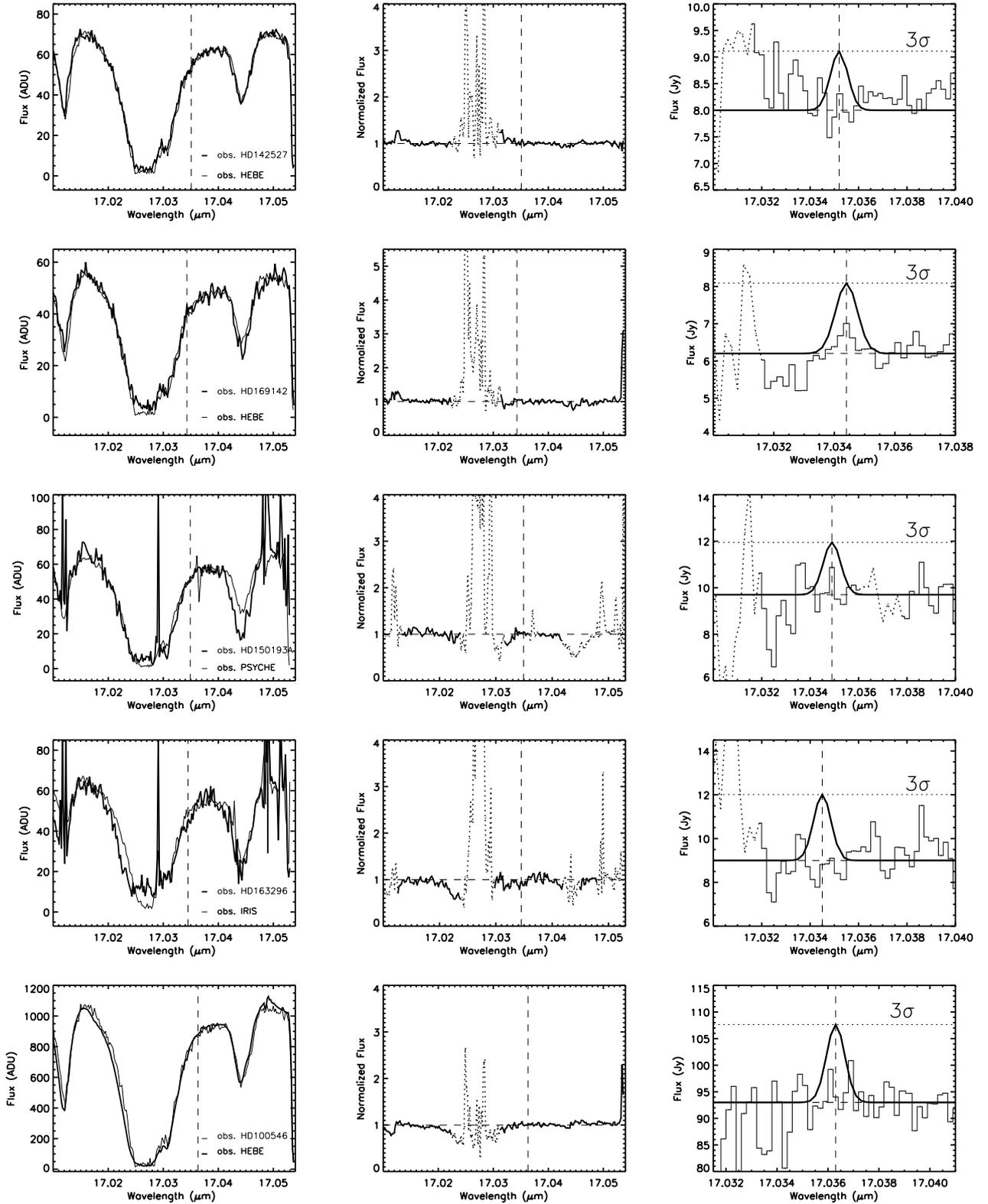}
\caption{Spectra obtained for the H$_2$ S(1) line at 17.035
  $\mu$m. {\it Left panel:} continuum spectra of the asteroid and of
  the target before telluric correction. {\it Middle panel:} full
  corrected spectra: dotted lines show spectral regions strongly
  affected by telluric features. {\it Right panel:} zoom of the region
  where the H$_2$ lines should be observed (dashed vertical lines).  A
  Gaussian of width FWHM = 21 \kms\ and integrated line flux equal to
  the 3$\sigma$ line-flux upper limits is overplotted.  The spectra
  were corrected neither for the radial velocity of the targets nor
  the Earth's rotation velocity.\vspace{3cm}}
\label{spectres}
\end{center}
\end{figure*}


\subsection{\visir\ observations}

We selected a sample of five Herbig Ae/Be stars with spatially
resolved disks (Table~\ref{param}). They were selected using the
following criteria: {\it (i)} proximity to the Sun ($<$150pc), with
well known distances; {\it (ii)} resolved disk and/or firm evidence of
gas-rich circumstellar environment, and/or properties close to that of
HD~97048 and/or AB Aur, for which the 17.035 $\mu$m line has already
been detected; {\it (iii)} well known Herbig Ae/Be stars with a rich
set of existing observations, particularly including gas-lines
spectroscopy observations \citep[Far-UV H$_2$ lines, CO lines...;
e.g. ][]{Thi01, klr08a, VdPlas_09} and PAHs (Polycyclic Aromatic
Hydrocarbons) features \citep[e.g. ][]{Habart04a, Acke06b}. The
fundamental parameters of each target are presented in
Table~\ref{param}.

The target stars were observed on 2009 June 14 and 15, with the high
spectral resolution long-slit mode of the \visir\ \citep[{\it ESO VLT
  Imager and Spectrometer for the mid-InfraRed}][]{Lagage04}
instrument at the VLT. We focussed on the observations of the S(1)
pure rotational line of H$_2$ at 17.035 $\mu$m because it is the most
intense line observable from the ground, and because it is the only
line ever detected with \visir\ in disks of Herbig Ae/Be stars
\citep{klr07}. The central wavelength of the observations was thus set
to be 17.035 $\mu$m. We used the 0.75'' slit, providing a spectral
resolution of about 14\,000. The observation conditions are summarized
in Table~\ref{log_obs}. In order to correct the spectra from the
Earth's atmospheric absorption and to obtain the absolute flux
calibration, asteroids and standard stars were observed at nearly the
same airmass and seeing conditions as the objects, just before and
after observing the science targets. For this purpose, we also used
the modeled spectra of the standard stars \citep{Cohen99}, and those
of the asteroids \citep[for HEBE and IRIS see][and for PSYCHE, Mueller
2009, priv. comm.]{Mueller98, Mueller02}.  For details on the
observations and data reduction techniques see \cite{klr07} and
\cite{klr08b}.

\begin{table*}
\begin{center}
  \caption{Results of the H$_2$ S(1) line analysis.}
\begin{tabular}{lcccccccccccccccccc}
  \hline
  \hline   
  Star   & $\lambda _{obs}^{\rm (a)}$& Integrated flux$^{\rm (b)}$ & Intensity$^{\rm (b)}$                & \multicolumn{3}{c}{$N({\rm H}_2)$ upper limits$^{\rm (b)}$}   & &  \multicolumn{3}{c}{H$_2$ mass upper limits$^{\rm (b,c)}$}     \\
  HD     &  ($\mu$m)       & (\flux)                 & (ergs\,s$^{-1}$\,cm$^{-2}$\,sr$^{-1}$) &  \multicolumn{3}{c}{(cm$^{-2}$)}                  & &   \multicolumn{3}{c}{ ($M_{\rm Jup}$)} \\
  \cline{5-7}  \cline{9-11}
         &                 &                         &                                      &    150K  & 300K & 1000K                             & &     150 K &    300 K  &  1000 K \\
  \hline    
142527  & 17.0351          & $<$1.0$\times$10$^{-14}$ & $<$2.5$\times$10$^{-3}$ & 9.4$\times$10$^{22}$ & 6.1$\times$10$^{21}$ & 1.4$\times$10$^{21}$ & & 1.9$\times$10$^{-1}$ & 1.2$\times$10$^{-2}$ & 3.3$\times$10$^{-3}$ \\ 
169142  & 17.0343          & $<$1.8$\times$10$^{-14}$ & $<$4.2$\times$10$^{-3}$ & 1.6$\times$10$^{23}$ & 1.0$\times$10$^{22}$ & 2.4$\times$10$^{21}$ & & 3.5$\times$10$^{-1}$ & 2.1$\times$10$^{-2}$ & 6.1$\times$10$^{-3}$ \\ 
150193A & 17.0349          & $<$2.1$\times$10$^{-14}$ & $<$5.0$\times$10$^{-3}$ & 1.9$\times$10$^{23}$ & 1.2$\times$10$^{22}$ & 2.8$\times$10$^{21}$ & & 4.5$\times$10$^{-1}$ & 2.7$\times$10$^{-2}$ & 7.7$\times$10$^{-3}$ \\
163296  & 17.0345          & $<$2.8$\times$10$^{-14}$ & $<$6.7$\times$10$^{-3}$ & 2.5$\times$10$^{23}$ & 1.6$\times$10$^{22}$ & 3.8$\times$10$^{21}$ & & 3.9$\times$10$^{-1}$ & 2.4$\times$10$^{-2}$ & 6.9$\times$10$^{-3}$ \\
100546  & 17.0363          & $<$1.4$\times$10$^{-13}$ & $<$3.3$\times$10$^{-2}$ & 1.2$\times$10$^{24}$ & 7.9$\times$10$^{22}$ & 1.9$\times$10$^{22}$ & & 1.4                  & 8.4$\times$10$^{-2}$ & 2.4$\times$10$^{-2}$ \\  
  \hline
\end{tabular}
\begin{list}{}{} 
\item (a) expected position of the line in the observed spectra by
  correcting the radial velocity of the star for the epoch of
  observation.
\item (b) 3$\sigma$ upper confidence limits.
\item (c) masses of H$_2$ are calculated assuming the distances quoted
  in Table~\ref{param}. \\
\end{list}
\label{tab_flux}
\end{center}
\end{table*}

\subsection{\uves\ observations}

To better understand the detections of H$_2$ towards HD~97048 and
AB~Aur, we observed spectral lines of the CH and CH$^+$ molecules in
the optical range at high spectral resolution using \uves\ (ESO
program numbers 075.C-0637 and 078.C-0774). The blue arm of \uves\
(3700\AA\ - 5000\AA) gives acces to the electronic transitions of the
CH and CH$^+$ molecules that are good tracers of the H$_2$ formation
and excitation \citep{Federman82, Mattila86, Somerville89}. The
formation of CH is predicted to be controlled by gas-phase reactions
with H$_2$. CH is thus a good tracer of H$_2$ and their abundances are
generally strongly correlated. The formation of the CH$^+$ molecule
through the chemical reaction C$^+$ + H$_2$ needs a temperature of
about 4500~K to occur. Thus, the CH$^+$ molecule is a probe of hot and
excited media, which could be interpreted as material close to the
star, and this would allow us to better constrain the excitation of
H$_2$.

The spectra were reduced using the \uves\ pipeline v3.2.0
\citep{Ballester00} based on the ESO Common Library Pipeline, and were
corrected from the Earth's rotation in order to shift them in the
heliocentric rest frame. To analyze the absorption lines in the
observed spectra we used the \owens\ package \citep[for details,
see][]{LEMOIN02, klr08a}, which allows us to derive the characteristics
of the gaseous components, i.e.  column density N, radial velocity
\vrad\ and broadening parameter $b$.

\section{Results}
\label{results}

\subsection{\visir\ sample analysis}

We observed five well studied Herbig Ae/Be stars with the high
resolution spectroscopic mode of \visir\ to search for the H$_2$ pure
rotational S(1) line. None of the observed sources show evidence for
H$_2$ emission at 17.035 $\mu$m (Fig.~\ref{spectres}). In HD~169142,
the expected centroid for the H$_2$ line corresponds to an emission
feature in the spectrum. However, this feature has a
full-width-at-half maximum (FWHM) smaller than a spectral resolution
element ($\Delta v \sim$ 21 \kms), and an amplitude of about 1$\sigma$
that makes it insignificant in terms of detection \citep[for details
on the calculation of the standard deviation $\sigma$,
see][]{klr08b}. We did not consider it further as a detection. As
described previously, we insist that mid-IR H$_2$ lines only probe
warm gas located in the surface layers of the disks because of the
large opacity in the interior layers due to the dust. The surface
layers of the disks, from which solid-state emission features also
arise, surrounding our target stars therefore do not contain
sufficient warm gas to enable it to be detected in emission at mid-IR
wavelengths.
 
\begin{figure}[!htbp]
\begin{center}
\includegraphics[width=8cm]{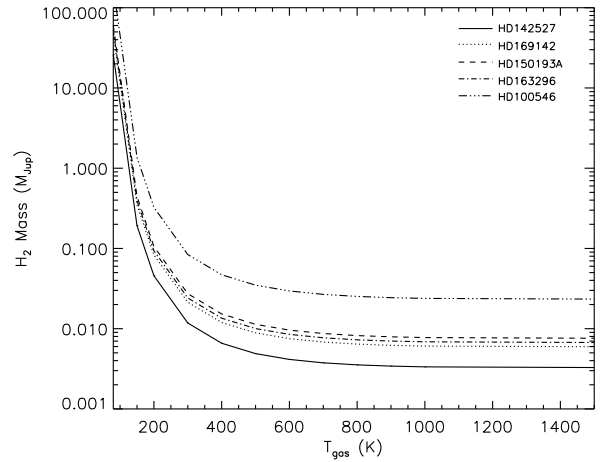}
\caption{Upper mass limits of optically thin H$_2$ derived from H$_2$
  S(1) line flux limits as a function of the assumed LTE temperature.}
\label{mass}
\end{center}
\end{figure}

Then, we derived 3$\sigma$ upper limits on the integrated line fluxes
and upper limits for the total column densities and masses of H$_2$ as
a function of the temperature for all the spectra by integrating over
a Gaussian of FWHM corresponding to the \visir\ spectral resolution,
and an amplitude of 3$\sigma$, centered on the expected wavelength for
the S(1) line (see Table~\ref{tab_flux} and Fig.~\ref{mass}).

For this purpose, we used the method detailed in \cite{klr08b,
  klr09a}, by assuming that the line is optically thin at local
thermodynamic equilibrium (LTE) and that the radiation is isotropic
\citep[for details on the formulae, see also][]{Van_Dishoeck_92,
  Takahashi01b}.

From the 3$\sigma$ upper limits to the emission line flux, we
calculated upper limits on the column density and mass of H$_2$ for
each star. We found that the column densities should be lower than
$\sim$10$^{24}$ cm$^{−2}$ at 150~K, and lower than $\sim$10$^{22}$
cm$^{−2}$ at 1000~K. The corresponding upper limits to the masses of
warm gas in the surface layers of the inner disk were estimated to be
in the range from $\sim$6$\times$10$^{-3}$ to $\sim$1.4 $M_{\rm Jup}$
(1 $M_{\rm Jup}$$\sim$10$^{-3}$~$M_{\odot}$), assuming LTE excitation,
and depending on the adopted temperature (see Table~\ref{tab_flux} and
Fig.~\ref{mass}).

\subsection{H$_2$ detection statistics}

In the past three years, numerous observations have been performed to
observe the pure rotational mid-IR emission lines of molecular
hydrogen in the circumstellar environments of HAeBes. \visir\
observations of 15 HAeBes have been conducted \citep[][and this
work]{Carmona08, klr07, klr08b} and only one source out of the 15,
namely HD~97048, presents a clear evidence for H$_2$ emission at
17.035 $\mu$m. In addition, in their sample of five HAeBes observed
with \texes\ at 17.035 $\mu$m, \cite{Bitner08} have reported only one
detection of the S(1) H$_2$ line in the disk of AB~Aur. This leads to
a detection statistics of the S(1) line from the ground of 10\% only
in disks about Herbig Ae/Be stars.  However, this statistic suffers
from the fact that the 3$\sigma$ limits in our sample are
inhomogeneous. Indeed, due to the different detection limits of each
instrument (and exposure time for each target), the 3$\sigma$ upper
limits on the line flux from \visir\ observations cannot be directly
compared to those obtained with \texes. However, our sample is quite
representative of the variety of HAeBes observable from the ground,
including stars with spectral type from F6 to B2, with very young,
massives or transitional disks. Therefore, the 10\% statistics can be
regarded as a lower limit on the number of H$_2$ detections with
current ground-based mid-IR instrumentation. This is consistent with
the models by \cite{Carmona08}, who show that mid-IR H$_2$ lines
cannot be detected with the existing instruments for disks under LTE
conditions. Those authors estimated that sensitivities down to
10$^{-16}$ \flux\ must be reached to detect the S(1) line, while
\visir\ typically reaches 10$^{-14}$ \flux.

The numerous non-detections of the S(1) line in circumstellar
environments of HAeBes could imply that the physical conditions of the
warm gas in most disks are consistent with those assumed in such a
model, i.e. gas and dust well-mixed, a gas-to-dust ratio of about 100,
and $T_{gas}=T_{dust}$.

\section{Discussion}
\label{discuss}

\subsection{H$_2$ and disk properties}

The challenge now is to explain the two detections of the S(1) line in
the disks of HD~97048 and AB~Aur. If the HD~97048 and AB~Aur disks
trace a particular evolutionary stage, one may expect some connection
between the mass of warm H$_2$ and the disk properties. 

The slope of the spectra corresponding to the ratio of the $F_{30}$
and $F_{13}$ indexes, i.e. continuum fluxes at 30 and 13 $\mu$m
respectively, has been used by \cite{Brown07} to identify the
population of cold disks.  In addition, young flared disks tend to
show rising spectra in the mid-IR with a much higher $F_{30}$ /
$F_{13}$ ratio than the more evolved, settled systems with
self-shadowed disks with much flatter mid-IR spectra
\citep{Dominik03}. The $F_{30}$ / $F_{13}$ ratio diagnostic is
therefore not unique, but allows us to identify a population with
properties that depart from those of most young stars with
disks. Indeed, in their sample of T Tauri stars, \cite{Brown07} have
shown that the majority of their sample had an emission that increased
by a factor of 2.3$\pm$1.4 between 13 and 30 $\mu$m, while the cold,
possibly transitional, disks rose by factors of $5-15$.

Here we measured the slope of the spectra of our targets, obtained
with the IRS spectrograph installed onboard the {\it Spitzer Space
  Telescope}, by computing $F_{13}$ which is the mean flux value (in
Jy) in the range 13 $\mu$m $\pm$ 0.5 $\mu$m, and $F_{30}$ the mean
flux value (in Jy) between 29 $\mu$m and 31 $\mu$m
\citep[Table~\ref{rapport_flux}, for details see][]{Olofsson09}. The
data reduction was performed with the ``c2d legacy team pipeline''
\citep{Lahuis2006} with the S18.7.0 pre-reduced (BCD) data.


\begin{table}
\begin{center}
  \caption{ $F_{30}$ / $F_{13}$ indexes, and H$_2$ mass upper limits
    at 600~K.}
\begin{tabular}{lccccccccccccccccc}
  \hline
  \hline 
  Star       & $F_{30}$ / $F_{13}$ & & H$_2$ mass and upper \\
             &                    & & limits at T=600~K ($M_{\rm Jup}$) \\
  \hline 
  (This work)&                    & &                     \\
  HD~142527  & 4.86               & & $<$4.1$\times$10$^{-3}$  \\
  HD~169142  & 7.12               & & $<$7.5$\times$10$^{-3}$  \\
  HD~150193A & 1.44               & & $<$9.6$\times$10$^{-3}$  \\
  HD~163296  & 1.54               & & $<$8.5$\times$10$^{-3}$  \\
  HD~100546  & 3.32               & & $<$2.9$\times$10$^{-2}$  \\
  \hline 
  \citep{klr08b}&                 & &                     \\
  HD~98922   & 0.70               & & $<$7.1$\times$10$^{-2}$  \\
  HD~250550  & 2.67               & & $<$1.2$\times$10$^{-1}$   \\
  HD~259431  & 2.78               & & $<$3.1$\times$10$^{-2}$   \\
  HD~45677   & $^{\rm (a)}$        & &                     \\
  \hline 
  \citep{Carmona08}&              & &                      \\
  UX~Ori     & $^{\rm (a)}$        & &                \\
  HD~100453  & 5.41               & & $<$3.8$\times$10$^{-3}$  \\   
  HD~101412  & 0.92               & & $<$7.3$\times$10$^{-3}$  \\
  HD~104237  & 1.28               & & $<$3.5$\times$10$^{-3}$  \\
  HD~142666  & 1.55               & & $<$4.7$\times$10$^{-3}$  \\
  \hline 
\citep{Bitner08}&               & &                     \\
49~Cet     & 2.31               & & $<$6.6$\times$10$^{-4}$  \\ 
MWC~758    & 4.00               & & $<$5.3$\times$10$^{-3}$ \\
V892~Tau   & 3.72               & & $<$1.9$\times$10$^{-2}$  \\
VV~Ser     & 0.76               & & $<$1.5$\times$10$^{-2}$  \\
\hline 
\citep{klr09a}&                  & &                               \\
HD~97048 $^{\rm (b)}$  & 6.22               & & (1.6$\pm$0.1)$\times$10$^{-2}$ \\
\hline 
\citep{Bitner07}&               & &             \\
AB~Aur $^{\rm (b)}$ & 4.21               & & (1.6$\pm$0.47)$\times$10$^{-3}$ \\
\hline 
\end{tabular}
\begin{list}{}{} 
\item (a) Spitzer data not available.
\item (b) masses of H$_2$ (not upper limits) because the S(1) has been
  detected.
\item We used distances and integrated fluxes values from the papers
  referenced in the table.
\end{list}
\label{rapport_flux}
\end{center}
\end{table}


\begin{figure}[!htbp]
\begin{center}
\includegraphics[width=8cm]{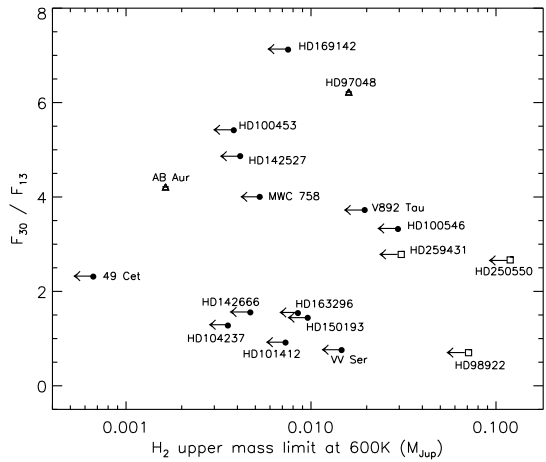}
\caption{Slope of the Spitzer spectra ($F_{30} / F_{13}$ ratio) versus
  upper mass limits of optically thin H$_2$ at 600~K derived from
  H$_2$ S(1) line flux limits. In this diagram are plotted all the
  Herbig Ae/Be stars observed at 17.035 $\mu$m with \texes\ and
  \visir. Triangles represent the two stars for which the S(1) line
  has been detected. Squares represent the Herbig Be stars for which a
  circumstellar disk has never been clearly detected. Plain circles
  represent the Ae stars of the sample (see text).}
\label{slope}
\end{center}
\end{figure}


Figure~\ref{slope} shows the upper limits of H$_2$ mass measured from
the S(1) line flux versus the slope of the spectra ($F_{30} / F_{13}$
ratio). For comparison with our targets, we plotted all the stars
observed at 17.035 $\mu$m by \visir\ and by \texes, including HD~97048
and AB~Aur \citep{klr07, klr08b, klr09a, Carmona08, Bitner08}. For
this task, we considered only the upper limits on the mass of H$_2$ at
600~K for each target, because the temperature of the observed H$_2$
is around 600~K for AB~Aur \citep{Bitner07} and lower than 600~K for
HD~97048 \citep{klr09a}.

Most of our sample stars have an emission that increases by a median
factor of 2.2$\pm$1.5 between 13 and 30 $\mu$m, which is fully
consistent with the results of \cite{Brown07}. However, four stars in
our sample, namely HD~169142, HD~100453, HD~142527, and HD~97048, have
significantly higher $F_{30} / F_{13}$ ratios. We stress that AB~Aur
is marginally higher than the bulk. In addition, the four spectra are
also characterized by the absence of 10 $\mu$m amorphous silicate
features. This is consistent with the fact that the strong wings of
the amorphous silicate feature at 10 $\mu$m increase the continuum
flux around 13$\mu$m and thus decrease the $F_{30} / F_{13}$ ratio.
The $F_{30} / F_{13}$ ratio of these four stars correspond to the
region of the plot where \cite{Brown07} have identified the
transitional cold disks in their sample, which also do not present the
10 $\mu$m amorphous silicate feature in their spectra. However,
HD~97048 is supposed to be a young star harboring a massive gas-rich
flared disk \citep{Lagage06, Doucet07}, as compared to the other three
stars that are known to have more evolved and less massive disks than
HD~97048 \citep[e.g.][]{Grady07}. Its position in our Fig.~\ref{slope}
raises numerous questions about its status. Below, we explore
different ways to understand the origin of the detected H$_2$ and the
status of HD~97048 (and marginally AB~Aur).

\subsection{Is the H$_2$ related to PAHs ?}

\begin{figure}[!htbp]
\begin{center}
\includegraphics[width=8cm]{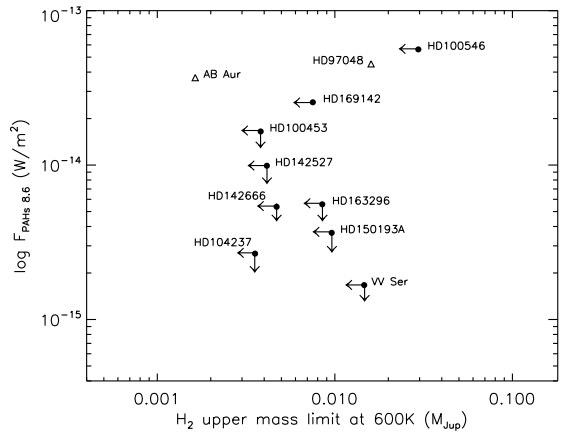}
\caption{Line flux of the PAHs feature at 8.6$\mu$m versus upper mass
  limits of optically thin H$_2$ at 600~K derived from H$_2$ S(1) line
  flux limits. Same legend as for Fig.~\ref{slope}.}
\label{PAH8}
\end{center}
\end{figure}

\begin{figure}[!htbp]
\begin{center}
\includegraphics[width=8cm]{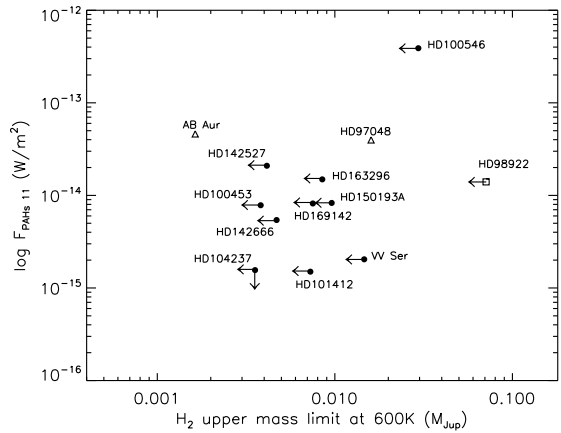}
\caption{Line flux of the PAHs complex at 11$\mu$m versus upper mass
  limits of optically thin H$_2$ at 600~K derived from H$_2$ S(1) line
  flux limits. Same legend as for Fig.~\ref{slope}.}
\label{PAH11}
\end{center}
\end{figure}

As we did for the $F_{30} / F_{13}$ ratio, we searched for the
possible correlation between the presence and line flux of PAHs
features and the upper limits on H$_2$ mid-IR lines. For this task we
compared the line flux of the PAHs feature at 8.6$\mu$m and 11.3$\mu$m
\citep[called 11$\mu$m complex due to the possible blend with features
of amorphous and crystalline silicate; e.g.][]{Acke04b} found in the
literature (Table~\ref{tab_PAHs}), with the H$_2$ upper mass limits
(Table~\ref{rapport_flux}). With regard to the $F_{30} / F_{13}$
ratio, AB~Aur and HD~97048 seem to depart from the bulk, and appear in
the top of the plots where the PAHs line fluxes are high
(Fig.~\ref{PAH8} and Fig.~\ref{PAH11}).

\begin{table}
\begin{center}
  \caption{Line fluxes of the PAHs features at 8.6$\mu$m and line
    fluxes of the PAHs complex at 11$\mu$m. }
\begin{tabular}{lrcrcc}
  \hline
  \hline
  star         & PAHs 8.6  (W/m$^2$)   & & Comp 11  (W/m$^2$)     & & ref   \\
  \hline
  HD~142527  & $<$9.90$\times$10$^{-15}$ & & 2.09$\times$10$^{-14}$   & & 1  \\
  HD~169142  & 2.55$\times$10$^{-14}$    & & 8.18$\times$10$^{-15}$   & & 1  \\
  HD~150193A & $<$3.64$\times$10$^{-15}$ & & 8.29$\times$10$^{-15}$   & & 1  \\
  HD~163296  & $<$5.57$\times$10$^{-15}$ & & 1.49$\times$10$^{-14}$   & & 1  \\
  HD~100546  & 5.61$\times$10$^{-14}$    & & 3.89$\times$10$^{-13}$   & & 1\\
  HD~98922   & y                        & & 1.4$\times$10$^{-14}$     & & 2, 3 \\
  HD~250550  &  n                       & &       n                  & & 4  \\
  HD~259431  & --                       & &       --                 & & --    \\
  HD~45677   & --                       & &       --                 & & --   \\
  UX~Ori     & $<$1.57$\times$10$^{-15}$ & & $<$7.20$\times$10$^{-16}$ & & 1   \\
  HD~100453  & $<$1.65$\times$10$^{-14}$ & & 7.83$\times$10$^{-15}$    & & 1\\   
  HD~101412  & y                        & & 1.5$\times$10$^{-15}$     & & 2, 3   \\
  HD~104237  & $<$2.67$\times$10$^{-15}$ & & $<$1.56$\times$10$^{-15}$ & & 1    \\
  HD~142666  & $<$5.37$\times$10$^{-15}$ & & 5.42$\times$10$^{-15}$    & & 1 \\
  49~Cet     & --                       & &       --                 & & --    \\ 
  MWC~758    & --                       & &       --                 & & --      \\
  V892~Tau   & --                       & &       --                 & & --   \\
  VV~Ser     & $<$1.67$\times$10$^{-15}$ & & 2.04$\times$10$^{-15}$   & & 1\\
  HD~97048   & 4.52$\times$10$^{-14}$    & & 3.97$\times$10$^{-14}$   & & 1\\
  AB~Aur     & 3.68$\times$10$^{-14}$    & & 4.61$\times$10$^{-14}$   & & 1 \\
  \hline 
\end{tabular}
\begin{list}{}{} 
\item -- Data not available.
\item ``Y'' (yes): presence of PAHs features but flux not measured,
\item  ``N'' (no): absence of PAHs features.
\item References: 1: \cite{Acke04b}; 2: \cite{Geers06}; 3:
  \cite{Geers07a}; 4: \cite{Habart04a}; \\
\end{list}
\label{tab_PAHs}
\end{center}
\end{table}

In the upper disk surface layers, photoelectric heating is very
efficient on small grains such as PAHs, and can play a significant
role in the gas heating process \citep{Kamp04, Jonkheid07}. The PAHs
emission depends on the geometry of the disk, i.e. the emission is
proportional to the amount of PAHs that is directly illuminated by the
UV light from the star. Thus the PAHs emission depends both on PAHs
abundance and on geometry, and it is unclear which of these two
effects is responsible for the higher PAH emission strength in AB~Aur
and HD~97048 compared to other Herbig stars in the sample.
Irrespective of its origin, the strong emission of PAHs appears
related to H$_2$ emission. Is there a link between their excitation
and heating mecanisms ? This question must be studied further.

As shown in Fig.~\ref{PAH8} and Fig.~\ref{PAH11}, the spectra of the
other stars of the sample present weaker PAH features. One could
assume that for these stars photoelectric heating is less efficient
and does not excite enough the H$_2$ to be detected in the mid-IR
range.

In both plots HD~100546 also departs from the bulk. This may be
because the disk of this star has a particular structure with an inner
ring, a gap between $\sim$0.5 and 10 AU, and an outer disk
\citep{Bouwman03, Benisty10}.


\subsection{Could the H$_2$ emission arise from the envelope? }

Both HD~97048 and AB~Aur are known to possess extended envelopes with
a significant contribution in the various observations of gas and dust
\citep[e.g.][]{Hartmann93, Grady99}. Could the observed H$_2$ mid-IR
emission be due to the gaseous component of the envelopes of these
stars?

This scenario can certainly excluded because, firstly, the H$_2$ S(1)
lines observed in the spectra of HD~97048 and AB~Aur are not spatially
resolved, which constrains the emitting region to be very close to the
central stars. Given the weakness of the detected H$_2$ emission, at
least for AB Aur, it is also unlikely that had any extended emission
is present. This would have been detected. Secondly, electronic
transitions of H$_2$ have been detected in absorption with the \fuse\
({\it Far Ultraviolet Spectroscopic Explorer}) satellite towards
AB~Aur, likely arising from the extended envelope surrounding the star
\citep{ROBERGE01, klr08a}. The column density of the $J=3$ level of
H$_2$ measured in the \fuse\ spectrum is 10$^5$ times lower than that
derived from the mid-IR observations, implying that we clearly do not
observe the same regions around the star.

\begin{figure*}[!htbp]
\begin{center}
\includegraphics[width=17cm]{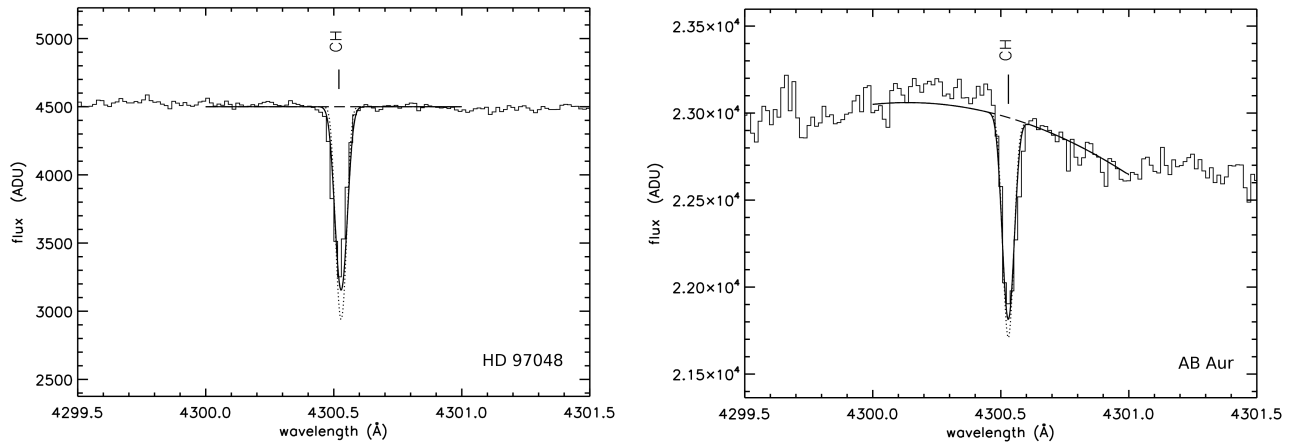}
\caption{Fits of the CH line at 4300.3~\AA\ in the \uves\ spectra of
  HD~97048 (left) and AB~Aur (right) with the \owens\ profile fitting
  procedure. Stellar continuum: dashed line; intrinsic line profile:
  dotted line; resulting profile convolved with the line-spread
  function: thick line.}
\label{fit_CH}
\end{center}
\end{figure*}

The source HD~97048 has never been observed with \fuse, which
precludes any other analysis of the H$_2$ lines than that of \visir\
observations. We thus analyzed the spectra of HD~97048 and AB~Aur
obtained with VLT/\uves.  For the two stars we detected absorption
lines of the CH and CH$^+$ molecules (see Fig.~\ref{fit_CH}), as well
as absorption lines of Ca~I, Ca~II and K~I. Due to the high
inclination angles of the disks with respect to the lines of sight,
the observed gas cannot be in the disks, but likely arises from the
surrounding envelope. We measured the radial velocities of these lines
in the \uves\ spectra of the two stars (Table~\ref{tab_obs_CH}). All
these lines in the \uves\ spectra have the same radial velocity,
implying that we probed a single gaseous component along each line of
sight. This is confirmed by the measured $b$-values of CH and CH$^+$
lines that are in the range expected for pure thermal broadening, as
generally seen in diffuse ISM clouds where typical $b$-values are
about 1.5 up to 3 \kms\ \citep[e.g.][]{Gry02}, and which would be much
higher if several gaseous components were present along the line of
sight. In addition, the radial velocities of the S(1) lines observed
for HD~97048 and AB~Aur, with \visir\ and \texes\ respectively
\citep{klr07, Bitner07} differ by about 10 \kms\ from those measured
for the gas in the \uves\ spectra. This is an additional clue that we
probed different regions of the environment of the stars in the mid-IR
and visible ranges, and that the H$_2$ emission in the mid-IR does not
come from the envelope, but from the disk.

Finally, it is important to note here that in their sample of Herbig
Ae/Be stars observed both with \fuse\ and \uves,
\cite{klr09_sf2a_talk, klr_CH} showed that when H$_2$ and CH are both
observed, their column densities are correlated in proportions
consistent with those observed in the diffuse interstellar medium
\citep[N(CH)/N(H$_2$)$\sim$10$^{-7}$, e.g.][]{Welty06, Sheffer08}. It
is important to note that for these stars, including AB~Aur, the
\fuse\ H$_2$ data implied an interstellar origin for the gas or a
remnant of the parent molecular cloud and/or circumstellar envelope
\citep{klr08a}. Here, the total column densities of H$_2$ derived from
the mid-IR observations give a N(CH)/N(H$_2$) ratio higher by a factor
10$^2$ to 10$^3$ than the interstellar ratio, again suggesting a
different location, the H$_2$ detected in the mid-IR being in the
disk, CH being in the envelope. Cold molecular hydrogen is very likely
present in the envelope as probed by \fuse, but does not possess the
physical properties required to emit at mid-IR wavelengths.

\begin{table}
\begin{center}
  \caption{Column densities, radial velocities and $b$ parameters of
    CH and CH$^+$ for HD~97048 and AB~Aur.}
\begin{tabular}{lcccccccccccccccccc}
  \hline
  \hline
  Star    &  log N(CH)   & log N(CH$^+$) &   \vrad      &  $b$      \\
          & (cm$^{-2}$)   & (cm$^{-2}$)   & (\kms)       & (\kms)      \\
  \hline
 HD~97048  & 13.1$\pm$0.4 & 12.2$\pm$0.3 & 14.5$\pm$0.2 & 1.5$\pm$0.4 \\
 AB Aur    & 12.3$\pm$1.4 & 13.1$\pm$0.1 & 15.0$\pm$0.1 & 1.8$\pm$0.3   \\
\hline
\end{tabular}
\begin{list}{}{} 
\item The radial velocities are in the heliocentric rest frame.
\end{list}
\label{tab_obs_CH}
\end{center}
\end{table}

\subsection{Comparison with CO observations} 

An important tracer of gas in the inner disks of T Tauri and Herbig
Ae/Be stars are the fundamental CO ro-vibrational emission bands
(e.g., $\Delta$v=1) present at 4.7$\mu$m. CO emission at 4.7 $\mu$m
traces warm and hot gas at temperatures ranging from hundreds to a few
thousands Kelvin \citep{Najita03}. For Herbig Ae stars, CO has been
detected in almost all sources with optically thick disks
\citep[i.e. E(K-L) $>$ 1][]{Brittain07}, typically exhibiting
rotational temperatures between 1000-1500~K and in a few cases hotter
than 2000~K \citep{Blake04, Brittain07}.  In Herbig Ae stars, the CO
emission properties and excitation mechanism appear to depend on the
disk geometry deduced from the SED shape \citep{Meeus01,
  DULLEMOND04}. The dominant excitation mechanism appears to be LTE
for self-shadowed disks, and fluorescence for flaring disks (van der
Plas et al. 2010, in prep). High-excitation CO emission lines (up to
v=5-4) have been observed only in flared disks (van der Plas et
al. 2010, in prep).

In the context of H$_2$ mid-IR emission detections, we find that in
all the Herbig Ae/Be stars where H$_2$ mid-IR emission has been
detected their disks are flared, and CO emission at 4.7$\mu$m has been
detected too. Note however that the opposite is not true, as this and
other studies show, not all the sources with CO $4.7\mu$m emission
display H$_2$ mid-IR emission. \cite{Bitner08} show that when H$_2$
mid-IR is detected, the gas temperatures are significantly higher than
the equilibrium temperatures expected for the emitting regions, thus
suggesting that the gas temperature is higher than the dust and that
processes such as UV-X-rays or accretion heating may be important
\citep[a conclusion in agreement with the conclusions of][when
explaining non-detections of H$_2$ emission]{Carmona08}.

If we extend the idea of an additional source of heating to collisions
\citep{Bitner08, Carmona08} to analyze the CO and H$_2$ mid-IR
detections together (note that mid-IR H$_2$ traces warm gas and CO 4.7
$\mu$m hotter gas), we may expect that the sources with detected H$_2$
mid-IR emission would have CO emission at temperatures higher than the
equilibrium temperature for the respective emitting regions. In other
words, we will expect for sources with H$_2$ mid-IR emission to
observe CO lines at high excitation transitions. This argumentation
works at least for HD~97048, where CO emission up to v=5-4 has been
detected \citep{VdPlas_09}. Indeed, \cite{VdPlas_09} modeled the
$^{12}$CO emission at 4.7$\mu$m around HD~97048, observed with the
VLT/CRIRES, with a homogeneous Keplerian disk. Their modeling of CO
emission yields an inner radius of 11~AU.  Those authors concluded a
depletion of CO in the inner region of the disk, probably due to
photodissociation. Our previous estimate of the location of the H$_2$
mid-IR emission ($>$5~AU to 35~AU) is compatible with the CO inner
radius, but CO also traces gas in the disk at much larger radii (up to
100~AU). However, this line of reasoning cannot explain the case of
HD~100546, a star where van der Plas et al. (2010, in prep) detected
CO emission up to v=5-4, but we do not see H$_2$ mid-IR emission with
VISIR. But then, there is a large gap (r$\sim$14~AU) in the dust disk
of HD~100546, with very little material inside \citep{Bouwman03,
  Benisty10}. This is likely different from HD~97048 and AB~Aur.

In addition, \cite{Brittain03} conducted observations toward AB~Aur at
4.7$\mu$m of $^{12}$CO (v=1-0) emission using the NASA Infrared
Telescope Facility (IRTF) and the Keck Observatory. The excitation
diagram of CO (assuming LTE gas) gives a temperature of 70~K for the
cool gas corresponding to the low-J levels, and an excitation
temperature of 1540~K for the hot gas (high-J levels). This diagram is
interpreted by those authors as hot CO emission coming from the inner
rim of the disk ($\sim$1~AU from the star), while the cool emission
originates in the outer flared part of the disk ($>$ 8~AU from the
star). The excitation temperature and radial velocity found for the
cold CO are fully consistent with those found by \cite{klr08a} for the
cold H$_2$ observed in absorption by \fuse, as well as for the CH
observed with \uves. On the other hand, \cite{Bitner07} showed that
the H$_2$ emission at 17$\mu$m in the disk of AB~Aur arised around
18~AU from the central star, with an excitation temperature of about
670~K. Their derived gas temperature and distance from the central
star fall between the hot and cold components seen in the CO
observations \citep{Brittain03}, suggesting that the H$_2$ and CO
IR-emission does not come from the same region of the disk.

It is hard to drive definitive conclusions based on very few
detections. Further searches of high-excitation CO emission using
high-spectral resolution in sources with H$_2$ mid-IR detections (most
notably AB Aur) and further searches of H$_2$ mid-IR emission in
sources with high-excitation CO will be required to test the idea
whether H$_2$ mid-IR emission and high-excitation CO emission are
correlated. For instance, the main conclusion that we can arrive at is
that Herbig Ae stars appear to be in general a uniform group
concerning CO $\mu$m emission and H$_2$ emission in the sense that CO
is detected in optically thick disks and that H$_2$ mid-IR emission is
very weak (i.e. not detected), except for those sources with flaring
disks where an additional mechanism heats the H$_2$ to a detectable
level. The use of full disks models \citep[coupling gas and
dust;][]{Woitke10} could help us to understand the possible
correlation between CO and H$_2$, but this is beyond the scope of this
paper.


\section{Conclusion}

We reported here on a search for the H$_2$ S(1) emission line at
17.035 $\mu$m in the circumstellar environments of five well known
HAeBes with the high resolution spectroscopic mode of \visir. No
source shows evidence for H$_2$ emission at 17.035 $\mu$m. From the
3$\sigma$ upper limits on the integrated line fluxes, we derived
limits on column densities and masses of warm gas as a function of the
temperature. The present work brings to 18 the number of
non-detections of the H$_2$ S(1) line in a global sample of 20 Herbig
Ae/Be stars observed with \visir\ and \texes\ \citep{klr07, Bitner07,
  Carmona08, klr08b, Bitner08}.

The detections of H$_2$ emission at 17.035 $\mu$m by \cite{Bitner07}
and \cite{klr07} show that at least a few circumstellar disks have
sufficiently high H$_2$ mid-infrared emission to be observed from the
ground. The most likely explanation for this is that the optically
thin surface layers of the disk has $T_{gas} > T_{dust}$ and that the
gas-to-dust ratio is higher than the canonical ratio of 100
\citep{Carmona08, klr07}.  Indeed, in the surface layers of the disk,
low densities or dust settling and coagulation may conduct to a
spatial decoupling between the gas and the dust. Photoelectric heating
can thus play a significant role in the gas heating process and the
physical conditions may rapidly differ from the LTE ones, i.e.,
$T_{gas} > T_{dust}$ \citep{Kamp04, Jonkheid07}. On the other hand, UV
and X-ray heating can be responsible for the excitation of the
observed gas and can heat the gas to temperatures significantly hotter
than the dust \citep{Nomura05, Glassgold07, Ercolano08}. In any case,
one would need to observe the lower H$_2$ $J$-levels (i.e. $J=0$ and
$J=1$) to definitively constrain the kinetic temperature of the gas,
and better understand the excitation mechanisms responsible of the
mid-IR emission.

As a second step, we also performed a statistical analysis of the
whole sample of Herbig Ae/Be stars observed at 17.035 $\mu$m with
\visir\ as well as with \texes. This analysis allowed us to identify a
population of stars, including HD~97048 and marginally AB~Aur, with
properties that depart from the bulk of our sample. This raises
numerous questions about the origin of the detected gas and the status
of HD~97048 and AB~Aur. From our \uves\ observations we clearly
demonstrated that the observed mid-IR H$_2$ emission does not come
from the envelope, but from the disk.

Are HD~97048 and AB~Aur peculiar stars (and why)? Due to the
similarities of these two stars ($T_{eff}$, age, mass, disk size...),
one would expect that the physical conditions of their circumstellar
gas are typical of a particular (short) stage of evolution of the
disks. However, to confirm this assumption, we would need to observe
other similar HAeBes. Unfortunately, no other nearby Herbig Ae/Be star
observable with the existing instruments presents the same
observational properties. A global diagnostic of the gaseous content
of the disks is thus now required. To better constrain the physical
and chemical properties of the gas, multi-wavelengths observations and
a deep modeling would be very useful. In this context, the {\it
  Herschel} satellite will be very helpful because it will allow us to
constrain the thermodynamics in young disks by observing the gas and
the dust simultaneously.


\begin{acknowledgements}
  This work is based on observations obtained at ESO/VLT (Paranal)
  with \visir, program number 083.C-0910, and with \uves\ program
  numbers 075.C-0637 and 078.C-0774.  We warmly thank G. van der Plas
  for fruitful discussions about CO observations. C.M-Z. warmly thanks
  A. Smette (ESO) for discussions about \visir\ observations.  We
  thank ANR for financial support through contract ANR-07-BLAN-0221
  (Dusty Disks). A.C acknowledge support from a Swiss National Science
  Foundation grant (PP002--110504).  C.M-Z. was supported by a CNES
  fellowship.
\end{acknowledgements}


\bibliographystyle{aa}
\bibliography{haebe}

\end{document}